\documentclass{aa}

\usepackage{float}

\usepackage{graphicx}

\usepackage{txfonts}

\usepackage{xcolor}			

\usepackage{gensymb}

\usepackage[
colorlinks = true,
linkcolor = black,
urlcolor  = cyan,
citecolor = black,
anchorcolor = cyan
]{hyperref}

\begin{document}

\title{An astrometric mass estimate for asteroid (223) Rosa}

\author{M. Kretlow }
\institute{Instituto de Astrofísica de Andalucía, IAA-CSIC, Glorieta de la Astronomía s/n, 18008 Granada, Spain. \email{kretlow@iaa.es}  }

\date{}


\abstract
{Outer main belt asteroid (223) Rosa is a possible flyby target of opportunity for ESA's (European Space Agency) JUpiter ICy moons Explorer (JUICE) mission when passing the asteroid belt on the way to Jupiter. The very low albedo and the featureless red spectra indicate a P-type asteroid in the Tholen taxonomy, though the yet known bulk density did not match very well this classification. \\[-0.75em]}
%
{Aim of this work was to derive new estimates for the mass and for the bulk density for (223) Rosa.\\[-0.75em]}
%
{The mass of Rosa was derived by analyzing the gravitational deflection of small `test' asteroids which had a close encounter with Rosa in the past. To find such events suitable for the mass determination, an encounter search with about 900,000 asteroids over the time span $1980-2030$ was performed.\\[-0.75em]}
%
{Three encounters were identified from which two independent mass estimates for Rosa were derived: $M = (5.32 \pm 2.17) \times 10^{17}$~kg and $M = (3.15 \pm 1.14) \times 10^{17}$~kg, respectively. The weighted mean is $M = (3.62 \pm 1.25) \times 10^{17}$~kg. This yields to a bulk density of $\rho = 1.2 \pm 0.5 \mathrm{\,g\,cm^{-3}}$, when adopting an effective diameter of $D = 83 \pm 8$~km. This bulk density estimate is consistent with typical densities for Tholen taxonomy P-type asteroids.\\[-0.75em]}
%
{}


\maketitle

\section{Introduction}\label{sec:intro}
Asteroid (223)~Rosa (hereafter alternatively also just Rosa) has been recently proposed as potential ESA JUICE mission\footnote{Currently planned for launch in April 2023} flyby target of opportunity on its way to Jupiter~\citep{avdellidou_characterisation_2021,agostini_analysis_2022}. Rosa is a very dark ($p_V < 0.05$) outer main belt asteroid (proper semi-major axis $a_p = 3.09011$~au)\footnote{Proper elements taken from \href{https://newton.spacedys.com/astdys/}{AstDyS}}, with an effective diameter $D$ of about 83~km (see section \ref{sec:diam}), and a yet not fully understood composition (e.g. the taxonomic class) and evolution (e.g. the dynamical and collisional history). 

\smallskip
Important physical parameters characterizing an asteroid are the size, the mass, and therefore the bulk density. While diameter estimates do usually exist for an asteroid provided by at least one of various well known methods like radiometry, stellar occultations and (occultation) scaled lightcurve inversion (3-d models), mass estimates are much harder to derive and are known just for a small number (some $10^2$) of the known asteroids. 
Compilation of asteroid masses are given in the literature \citep[e.g.][and references therein]{hilton_asteroid_2002,carry_density_2012}, and more recently also provided by web based data archives \citep{kretlow_size_2020}.

Gravitational deflection of a significant smaller object (`test asteroid') by another asteroid (`perturber'), which mass is aimed to be derived during a mutual encounter, is one of the commonly applied methods in this domain \citep{hilton_asteroid_2002}. Beside such investigations on individual objects, asteroid masses are also derived as model parameter fit results of major planetary ephemeridies construction (simultaneously with other asteroid masses which are included in the dynamical model). 
Two mass estimates for (223)~Rosa were found in the literature, and both were derived in that way. \citet{fienga_inpop19a_2019} obtained a mass of $M = (5.979 \pm 2.971)\times 10^{17}$ kg. \citet{park_jpl_2021} provides a value of $M = 9.350 \times 10^{17}$ kg.

As mentioned diameter estimates are available for the majority of numbered asteroids, as they were observed by dedicated space-based infrared surveys like IRAS, WISE/NEOWISE and Akari and/or were obtained by other methods. Therefore in many cases more than one size estimate is found in the literature. Using an uncertainty-weighted average of $D = 82.7 \pm 8.4$~km for the mean diameter and a mass of $M = (5.979 \pm 2.971)\times 10^{17}$ kg from \citet{fienga_inpop19a_2019}, \citet{avdellidou_characterisation_2021} computed a bulk density of $\rho = 1.79 \mathrm{\,g\,cm^{-3}}$ with a relative uncertainty of about 50\%~\footnote{Note that I derive $\rho = 2.0 \pm 1.2  \mathrm{\,g\,cm^{-3}}$ for these values for $D$ and $M$}.

\smallskip

Section~\ref{sec:diam} gives an overview about available size measurements and aims to a (current) best value. In section~\ref{sec:mass} this work provides a new mass estimate derived from mutual encounters with other small asteroids. Section ~\ref{sec:results} summarizes the proposed best values for the size and mass of (223)~Rosa, provides the resulting bulk density and discusses these values with respect to the taxonomic class of the asteroid.

\section{Diameter estimates}\label{sec:diam}

The value (and uncertainty) of the diameter $D$ is crucial for the resulting bulk density due to $\rho \propto D^{-3}$. As mentioned in the introduction mostly thermal diameters were published so far. Table~\ref{tab:diameters} presents these literature values found up to now.
It is out of the scope and beyond the capability of this work to discuss and to evaluate every diameter estimate given in Tab.~\ref{tab:diameters} in order to identify a best value on an argumentative basis. Thus the effective diameter used in this work for Rosa is derived by averaging all these literature values.
Because in weighted arithmetic means the error $\sigma$ is used as weight $w=1/\sigma^{2}$, outliers with small formal uncertainties will bias the result. An alternative form of averaging is the expected value method (EVM, \citet{birch_method_2014}), which is more robust against outliers. Table~\ref{tab:diameters} provides both the traditional weighted average ($ D = 82.6 \pm 8.2$\,km) as well as the EVM average ($ D = 83.3 \pm 8.1$\,km) for the yet published diameter estimates.

\begin{table}
\caption{Compilation of literature values for the effective diameter, reproduced from \citet{avdellidou_characterisation_2021}  and completed with recent publications. Two different averages are also provided, the classical weighted (c.w.) mean and the EVM value. Entries with asymmetric uncertainties (entries \#13 and \#14, reference (j)) are therefore symmetrized~\citep[Appendix A, method 2]{audi_nubase2016_2017}.
References: (a) \citet{nugent_neowise_2016}, (b) \citet{masiero_neowise_2017}, (c) \citet{masiero_asteroid_2020}, (d) \citet{masiero_main-belt_2014}, (e) \citet{usui_asteroid_2011}, (f) \citet{masiero_main_2011}, (g) \citet{tedesco_supplemental_2002}, (h) \citet{ryan_rectified_2010}, (i) \citet{masiero_preliminary_2012}, (j) \citet{marciniak_properties_2021}. Methods: STM (standard thermal model), NEATM (near-Earth asteroid thermal model), CITPM (convex inversion thermophysical model). 
}
\label{tab:diameters}
\centering
\begin{tabular}{rcccc}
\hline\hline\\[-0.9em]
\# & Diameter $D$ (km)   &  Albedo $p_V$   &  Method   &  Ref. \\
\hline
 1 & 72.33 $\pm$ 20.18  & 0.040 $\pm$ 0.050 & NEATM & a \\
 2 & 76.46 $\pm$ 29.36  & 0.035 $\pm$ 0.023  & NEATM  & b \\
 3 & 79.68 $\pm$ 33.85  & 0.036 $\pm$ 0.037 &  NEATM  & c \\
 4 & 79.81 $\pm$ 0.31  & 0.040 $\pm$ 0.010  & NEATM &  d \\
 5 & 80.93 $\pm$ 1.46 &  0.037 $\pm$ 0.002 &  NEATM  & e \\
 6 & 83.39 $\pm$ 2.97  & 0.034 $\pm$ 0.005 &  NEATM  & f \\
 7 & 86.05 $\pm$ 26.43  & 0.033 $\pm$ 0.046 &  NEATM  &  c \\
 8 & 87.61 $\pm$ 4.40  & 0.031 $\pm$ 0.003  & STM  & g \\
 9 & 88.50 $\pm$ 3.79  & 0.031 $\pm$ 0.003  & NEATM  & h \\
10 & 89.37 $\pm$ 3.53  & 0.030 $\pm$ 0.002 &  STM &  h \\
11 & 90.43 $\pm$ 20.33  & 0.030 $\pm$ 0.020 &  NEATM &  a \\
12 & 109.16 $\pm$ 0.95  & 0.020 $\pm$ 0.003 &  NEATM &  i \\
13 & 72.8 $\pm$ 6.3  & 0.034 $\pm$ 0.005 &  CITPM &  j \\
14 & 72.5 $\pm$ 4.2  & 0.035 $\pm$ 0.005 &  CITPM &  j \\
[0.25em]
15 & 82.6 $\pm$ 8.2  & 0.032 $\pm$ 0.002 &  c.w. mean &  - \\
16 & 83.3 $\pm$ 8.1  & 0.033 $\pm$ 0.006 &  evm mean &  -\\
\hline
\end{tabular}
\end{table}

\begin{figure*}
\centering
\includegraphics[width=0.95\textwidth]{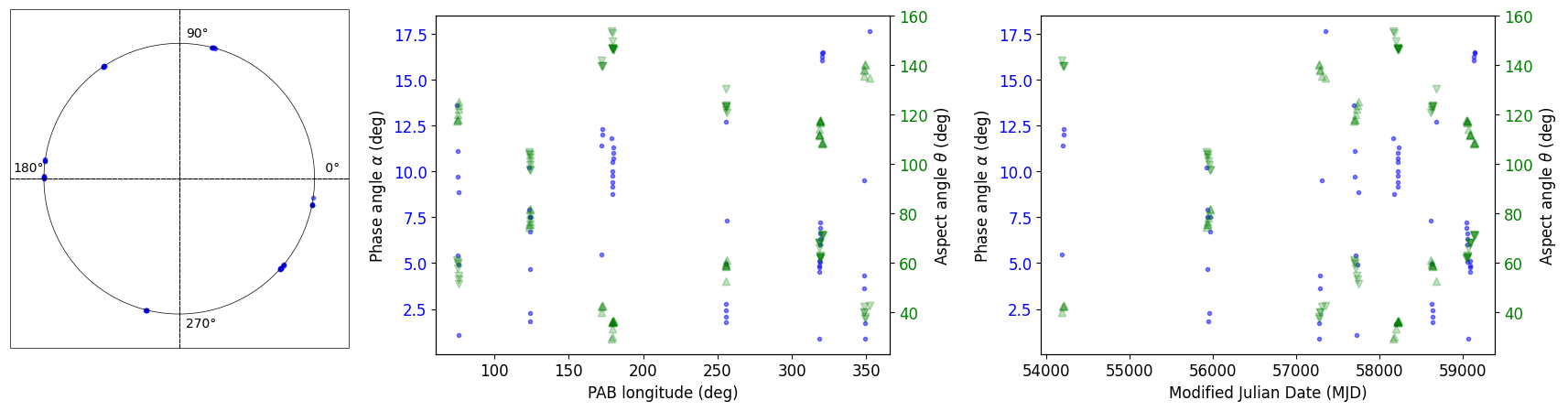}			
\caption{\small Aspect data for the lightcurves given in Fig.~\ref{fig:all_opt_lc}. The left panel shows the Phase Angle Bisector (PAB) longitude distribution. In the middle and right panel the phase and aspect angle $\alpha, \theta$ are given as function of PAB longitude (middle) and MJD (right). The phase angle $\alpha$ is marked by dots. The aspect angle $\theta$ is computed for both solutions derived by ~\citet{marciniak_properties_2021}. Solution $\lambda_p,\beta_p$ = ($22\degree,18\degree$) is marked by triangles ($\blacktriangle$), the mirror solution $\lambda_p,\beta_p$ = ($203\degree,26\degree$) by upside-down triangles ($\blacktriangledown$).}
\label{fig:aspect_data}
\end{figure*}

\smallskip
As shown in Fig.~\ref{fig:aspect_data} Rosa was observed in different viewing aspects, with phase angles $\alpha$ between $0.9\degree$ and $17.1\degree$ and aspect angles $\theta$ between $37.6\degree$ and $153.8\degree$ for pol solution 1 (respectively between $29.3\degree$ and $140.5\degree$ for pol solution 2) derived by \citet{marciniak_properties_2021}. The highest lightcurve amplitude in the set of archived lightcurves is Amax = 0.16~mag for a nearly equatorial view, which corresponds to an axes ratio $a/c = 1.16$ for an aspect angle $\theta = 90\degree$. The Asteroid Lightcurve Database (\href{https://www.minorplanet.info/PHP/lcdb.php}{LCDB}) provides values Amin = 0.06~mag and Amax = 0.13~mag. Putting this together we can expect that Rosa is neither very elongated nor irregular and the assumption of a spherical body made in simple thermal models like STM (standard thermal model) and NEATM (near-Earth asteroid thermal model) is appropriate for this asteroid. Averaging all available size estimates derived from STM, NEATM and also TPM (thermophysical model) as described beforehand seems reasonable in order to derive a reliable value for the effective diameter.

\smallskip 
The observation of stellar occultations by asteroids provides the shape and size of the apparent profile on the sky plane  with km accuracy. Up to now three stellar occultations by (223) Rosa were successful observed~\citep{herald_asteroid_2019}: 2004-09-29 (single visual observation with two subsequent occultation events; corresponding chord lengths 34~km and 78~km; but inconsistent with adjacent negative (miss) observations), 
2008-05-27 (2 chords), and 2014-04-13 (1 chord: length = 46.4~km).
The 2008 May 27 stellar occultation was recorded by three stations which were operated by the same observer (i.e. two unattended stations), yielding in two positive detections (chord lengths 54.5~km and 66.7~km) and one miss (no occultation detected).

\begin{figure*}
\centering
\includegraphics[width=1.0\textwidth]{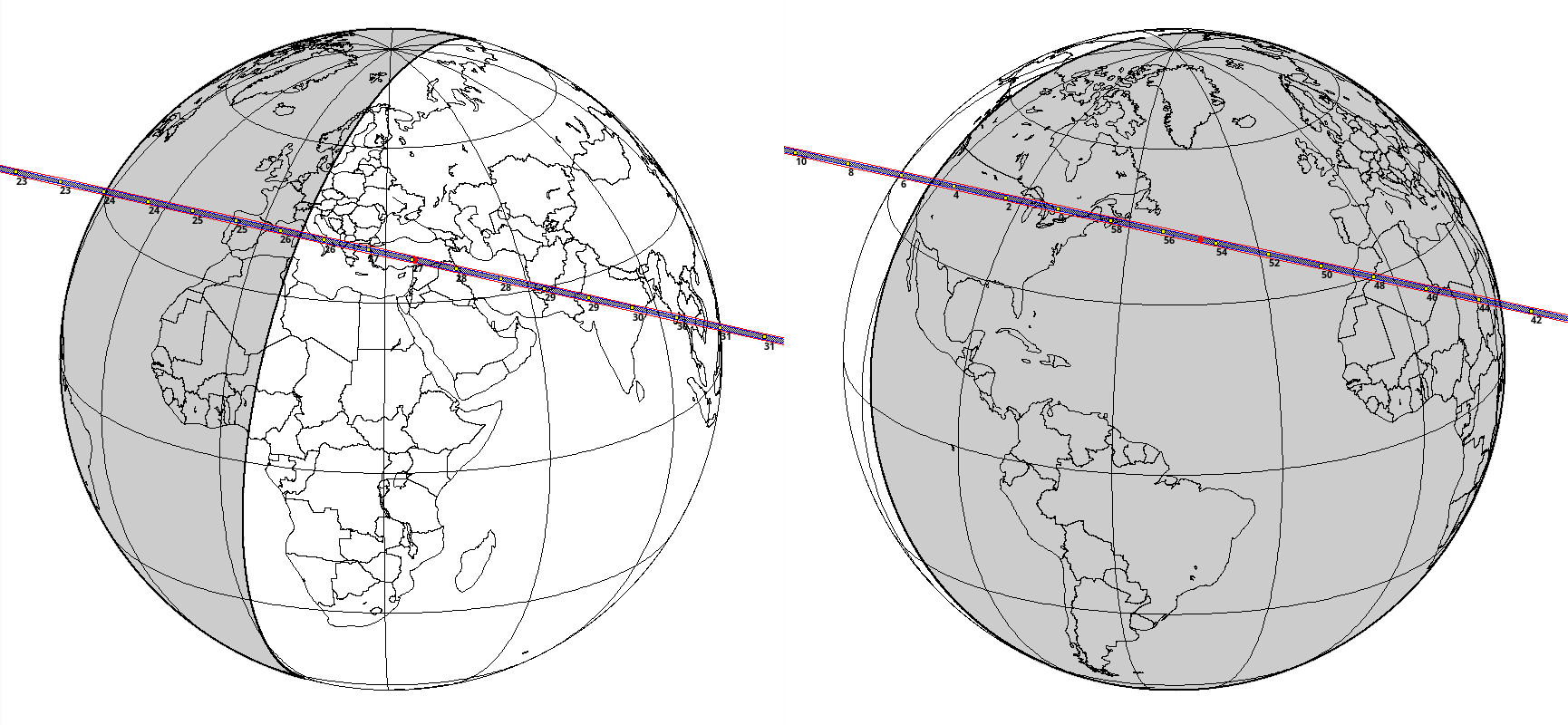}			
\caption{\small Two stellar occultation predictions for (223) Rosa. The red line marks the $1-\sigma$ cross-track uncertainty (about 24~km for both events). Minute time ticks are marked by yellow dots. The time of geocentric closest approach (c.a.) is marked by the red dot. \textsl{Left-hand side:} occultation of the star Gaia EDR3 0664382207483343488 on 2022-10-07 at 05:27:31 UT (c.a.), visible in Portugal and Spain (though, twilight will interfere in the eastern part of Spain).  The star has G = 13.5~mag, the expected magnitude drop will be 2.0~mag, the maximum duration about 2.9~sec. \textsl{Right-hand side:} occultation of the star Gaia EDR3 0661647103588763776 on 2023-02-24 at 01:54:39 UT (c.a.), visible in the U.S. and northern Africa (Morocco, Algeria, etc.) The star has G = 13.9~mag, the expected magnitude drop will be 0.8~mag, the maximum duration about 9.6~sec.}
\label{fig:occ_pred}
\end{figure*}

\smallskip
The observation of further stellar (multi-chord) occultation observations by Rosa will help to refine its volume, especially in combination with lightcurve data. Therefore stellar occultation predictions were performed until the year 2025, considering Gaia EDR3 \citep[Early Data Release 3,][]{brown_gaia_2021} stars down to G = 16 mag. These predictions will be available on the authors \href{https://astro.kretlow.de/?StOcc}{website}. Two upcoming events crossing over Europe (2022-10-07) and the USA (2023-02-24), regions which potentially have a good coverage by occultation observers, are presented in Figure~\ref{fig:occ_pred}.

\section{Mass estimates}\label{sec:mass}
Up to now two mass estimates were published in the literature from the construction of planetary ephemerides and only one of those includes error bars (see Sec.~\ref{sec:intro}). Aim of this work was to derive a new estimate for the mass of Rosa with the means of the gravitational deflection of small asteroids during close encounters.

\smallskip
In a first step a search for suitable test asteroids was performed essentially by integrating (223) Rosa and about  900,000 known asteroids\footnote{Orbital elements taken from MPC's \href{https://www.minorplanetcenter.net/iau/MPCORB.html}{MPCORB.DAT}} backwards until the year 1980 (and also into the future until the year 2030 in order to identify upcoming events). If the euclidean distance between Rosa and another (test) asteroid became smaller than 0.05 au during this ephemeris interval, the mutual encounter was stored together with additional information like the time span of available observations of the test asteroid, the relative encounter velocity, and the parameter $P = D_1^3/(r \, v)$; with $D_1$ the diameter of the perturber (set to $D_1 = 83 \pm 8 $\,km, see section~\ref{sec:diam}), $r$ the encounter distance between both objects in km and $v$ their relative encounter velocity in km/s.  $P$ is a proxy for the change in the mean motion $\Delta n$ of the perturbed body \citep{galad_asteroid_2001}. This resulting list of encounters was finally used to select candidates for the asteroid mass determination. For (223) Rosa three close encounters were identified to be potentially suitable for the mass determination (Table~\ref{tab:test_asteroids}). 

\smallskip
The mass $M$ of the perturbing body (Rosa) was determined by means of a Least-Squares-Fit of the solve-for parameters to the astrometric observations\footnote{Provided by the Minor Planet Center (\href{https://www.minorplanetcenter.net/}{MPC})} of the test asteroid by solving the system of linear equations
\begin{equation}
P \Delta E + Q \Delta M = R ,
\end{equation}
where $ P = \partial (\alpha,\delta)/\partial E$ is the matrix of partial derivatives of the observed coordinates $(\alpha,\delta)$ with respect to the six initial values $E_1,\dots,E_6$ (position and velocity) of the test asteroid. $Q =  \partial (\alpha,\delta)/\partial M$ is the matrix of partial derivatives of the observed coordinates of the test asteroid with respect to the perturbing mass $M$, and $R$ is the matrix depending on the (O-C) residuals in coordinates of the test asteroid. $\Delta E = (\Delta E_1,\dots,\Delta E_6)$ are the corrections to the six initial values of the test asteroid and $\Delta M$ is the correction to the mass of the perturbing body. The partial derivatives $P, Q$ were not computed by numerical variation, but rather by integrating a set of seven differential equations together with the equations of motion of the test asteroid \citep{sitarski_correction_1971}.

\smallskip
The numerical integration was carried out using 
a multi-step, variable order, predictor-corrector (PECE) method with self-adjusting step size \citep{shampine_computer_1975}. The masses of the planets and their state vector during the integration were read from an external file (JPL DE440). The perturbing asteroids of the dynamical model were handled in a similar way, as well as for the perturber Rosa. Ephemerides of all of them are precomputed and stored as Chebyshev polynomials in external files, to be used in the program on demand. The dynamical model is summarized in Table~\ref{tab:dyn_model}. If applicable, observations which have been reduced with non-Gaia star catalogues were de-biased \citep{eggl_star_2020}. For the weights of the observations during the differential orbit correction of the test asteroid the error estimates given by \cite{veres_statistical_2017} were used.

\begin{table}
\caption{List of the perturbing asteroids  included in the dynamical model. The masses of the asteroids are taken from \citet[\href{https://doi.org/10.5281/zenodo.4039774}{Data set}]{kretlow_size_2020}.}
\label{tab:dyn_model}
\centering
\begin{tabular}{lclc}
\hline\hline\\[-0.9em]
Asteroid     &  Mass (kg)   & Asteroid   &  Mass (kg) \\
\hline\\[-0.9em]
(1) Ceres    & $9.38\times 10^{20}$ & (16) Psyche     & $2.29\times 10^{19}$ \\
(2) Pallas   & $2.05\times 10^{20}$ & (29) Amphitrite & $1.33\times 10^{19}$ \\
(3) Juno     & $2.63\times 10^{19}$ & (52) Europa     & $2.34\times 10^{19}$ \\
(4) Vesta    & $2.59\times 10^{20}$ & (65) Cybele     & $1.63\times 10^{19}$ \\
(6) Hebe     & $1.28\times 10^{19}$ & (87) Sylvia     & $1.51\times 10^{19}$ \\
(7) Iris     & $1.41\times 10^{19}$ & (88) Thisbe     & $1.03\times 10^{19}$ \\
(10) Hygiea  & $8.42\times 10^{19}$ & (511) Davida    & $3.08\times 10^{19}$ \\
(15) Eunomia & $3.16\times 10^{19}$ & (704) Interamnia & $3.54\times 10^{19}$\\
\hline
\end{tabular}
\end{table}

\begin{table*}[th]
\caption{Test asteroids and their mutual encounter events used for the mass determination of (223) Rosa, ordered by encounter date. 
Given are the designation of the test asteroid, the date of the encounter, the minimum euclidean distance $r$ between Rosa and the test asteroid, the relative encounter velocity $v$, the diameter $D_2$ of the test asteroid (either taken from the orbital elements file \href{https://asteroid.lowell.edu/main/astorb/}{astorb.dat} or calculated from the absolute magnitude $H$), the perturbation parameter $P$, the time span covered by the astrometric observations filed at the \href{https://www.minorplanetcenter.net/}{MPC} (at date 2021-12-10), and the derived mass value $M_{223}$ for Rosa.} 
\label{tab:test_asteroids}
\centering
\begin{tabular}{ccccccccc}
\hline\hline\\[-0.9em]
Test asteroid & Encounter date &  $r$  & $v$ & $D_2$ & $P$  & Obs arc  &  $M_{223}$\\
              &                &       au                           & $\text{km s}^{-1}$   &  km &  km s &   yyyy.mm & $10^{17}$ kg\\ 
\hline
(78824) 2003 QS13   & 2001-06-26.93 &  0.0002 & 2.837 &  1.8 & $8.1 \pm 2.3$   &   1960.09 -- 2021.12 & --- \\
(35525) 1998 FV64    &  2010-12-31.87 &  0.0004 & 2.043 &  3.9 & $5.3 \pm 1.5$   &  1990.09 -- 2021.11 & $5.32 \pm 2.17$\\ 
(315162) 2007 FL24 & 2016-07-04.05 & 0.0027 & 0.131&  1.8 &  $12.3 \pm 3.5$  & 1999.11 -- 2021.11 &  $3.15 \pm 1.14$ \\
\hline
\end{tabular}
\end{table*}

\subsubsection*{(35525) 1998 FV64}
This encounter between 1998 FV64 and Rosa at the end of the year 2010 has the smallest perturbation parameter value of all three events presented here. The pre-encounter as well as post-encounter observation coverage is extensive and homogeneous. A mass estimate $M = (5.32 \pm 2.17) \times 10^{17}$ kg was derived for Rosa. 

\subsubsection*{(315162) 2007 FL24}
2007 FL24 had a long and slow encounter with Rosa in 2016 (thus having the largest perturbation parameter $P$ of all three cases), and also a good astrometric coverage both pre-encounter as well as post-encounter. The orbit solution for this encounter yields to a mass estimate $M = (3.15 \pm 1.14) \times 10^{17}$ kg.

\smallskip

Because the osculating orbital elements of Rosa and 2007 FL24 look very similar, the possibility of being an asteroid pair was investigated. First the distance between them in the 3-parameter space was calculated using their proper elements\footnote{Provided by \href{https://newton.spacedys.com/astdys/}{AstDyS}} by the formula \citep{zappala_asteroid_1990}:
\begin{equation}
d = n a \sqrt{k_a(\Delta a/a)^2 + k_e(\Delta e)^2 + k_i(\Delta \sin i)^2},
\end{equation}
with $a$ the mean of both semi-major axes, $n$ the corresponding mean motion, and the standard metric %
weighting factors $k_a = 5/4, k_e=k_i=2$. Though the high value of $d = 51 $ m/s made it unlikely that these two bodies are an asteroid pair, a 500\,kyr backward integration was made using the mercury6 software \citep{chambers_mercury_2012} regarding all major planets (M-N) plus the dwarf planet/asteroids Ceres, Pallas, Vesta, but without applying any non-gravitational forces. No close encounter / coincidence point was found in this time span, thus the orbital similarity seems to be a random coincidence.

\subsubsection*{(78824) 2003 QS13}
Though this encounter in 2001 has a perturbation parameter $P = 8.1 \pm 2.3$\, km\,s, which is even larger than that one for the case 1998 FV64, the orbit solution did not yield to a physical reasonable mass solution for Rosa. A likely explanation is that there are only four pre-encounter nights of observation, one night in September 1960 (Palomar Mountain, 675), one single observation in 1994 (Tautenburg, 033), and two sequential nights in December 1996 (Prescott, 684). A re-reduction of these pre-encounter observations (if available) using the Gaia star catalog could be helpful. A search in image archives\footnote{Using the CADC Solar System Object Image Search (\href{https://www.cadc-ccda.hia-iha.nrc-cnrc.gc.ca/en/ssois/index.html}{SSOIS})} with the aim to retrieve additional and precise positions was not succesfull, because the object was usually too faint to be detected on older survey images like NEAT (Near-Earth Asteroid Tracking) GEODSS/Maui. Table~\ref{tab:archive_images} lists the images which were examined for this asteroid. It is also conceivable that a different numerical approach, like for example a Monte Carlo parameter estimation \citep{siltala_asteroid_2020}, might be able to use this encounter for a mass determination with the existing astrometry. 

\smallskip
A summary of all three encounter circumstances and the derived mass results is given in Table~\ref{tab:test_asteroids}.

\section{Results and Discussion}\label{sec:results}

\begin{figure}
\centering
\includegraphics[width=1.0\columnwidth]{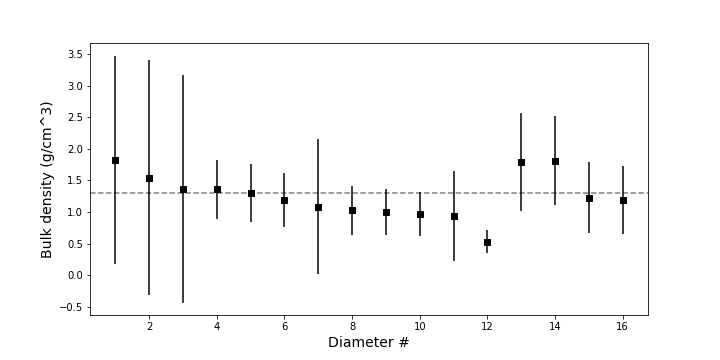}			
\caption{\small Bulk density as function of the diameters given in Table~\ref{tab:diameters} for the final mass estimate $M = (3.62 \pm 1.25) \times 10^{17}$ kg. The horizontal dashed line marks the typical density  $\rho \approx{}1.3 \mathrm{\,g\,cm^{-3}}$ for Tholen taxonomy P-type asteroids.}
\label{fig:bd_vs_diam}
\end{figure}

From two individual close encounters with small asteroids, the mass of Rosa was estimated to $M = (5.32 \pm 2.17) \times 10^{17}$ kg and $M = (3.15 \pm 1.14) \times 10^{17}$ kg, respectively. The weighted mean value is $M = (3.62 \pm 1.25) \times 10^{17}$ kg. Figure~\ref{fig:bd_vs_diam} shows the bulk density as function of the diameters given in Table~\ref{tab:diameters}. 
By adopting $D = 83.3 \pm 8.1$ km as (currently) best value for the diameter, the corresponding bulk densities are $\rho = 1.8 \pm 0.9 \mathrm{\,g\,cm^{-3}}$ and $\rho = 1.0\pm 0.5 \mathrm{\,g\,cm^{-3}}$. The weighted mean is $\rho = 1.2 \pm 0.5 \mathrm{\,g\,cm^{-3}}$. The EVM mean of the visual albedo is $p_V = 0.033 \pm 0.006$.

The mass values derived in this work are smaller than the values $M = (5.979 \pm 2.971)\times 10^{17}$ kg and $M = 9.350 \times 10^{17}$ kg \citep{fienga_inpop19a_2019,park_jpl_2021}, which yield to bulk densities of $\rho = 2.0 \pm 1.1 \mathrm{\,g\,cm^{-3}}$ and $\rho = 3.1 \mathrm{\,g\,cm^{-3}}$, respectively. 

From the reassessment of the existing spectra and spectrophotometric data, supplemented with own NIR and VIS spectra taken in 2021 and the very low albedo, \citet{avdellidou_characterisation_2021} concluded that Rosa is probably a Tholen taxonomy P-type asteroid, though the previously mentioned densities of $2.0 \pm 1.1 \mathrm{\,g\,cm^{-3}}$ and $3.1 \mathrm{\,g\,cm^{-3}}$ would be somewhat to large for this taxonomic class. 

\smallskip

However, Rosa's bulk density obtained in this work agrees well with typical densities of P-type asteroids like (87) Sylvia and (107) Camilla ($\rho \approx{}1.3 \mathrm{\,g\,cm^{-3}}$; see e.g. \citet{carry_evidence_2021}, \citet{vernazza_vltsphere_2021}).

\begin{figure*}[th]
\centering
\includegraphics[width=1.0\textwidth]{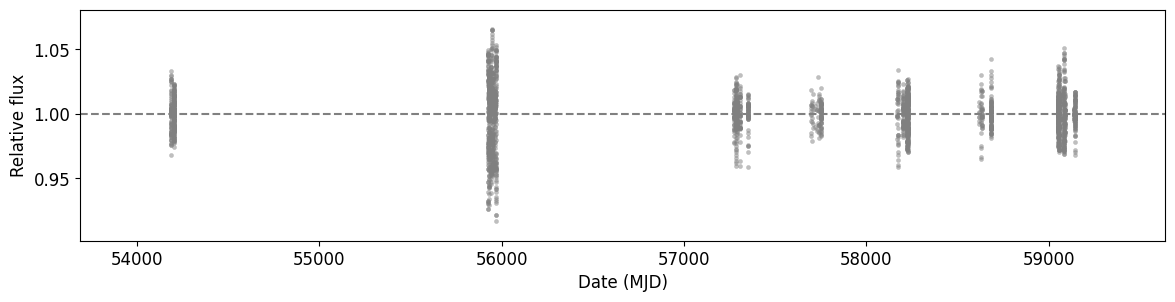}			
\caption{\small Overview about all 58 currently available optical lightcurves, archived at the \href{http://cdsarc.u-strasbg.fr/viz-bin/cat/J/A+A/654/A87}{CDS}~\cite[]{marciniak_properties_2021}. Given is the relative brightness in intensity units, normalized to the mean brightness vs time (Modified Julian Date, MJD) for the time span $2007-2020$. The maximum amplitude Amax for the epoch around MJD 55950 is 0.16~mag and corresponds to a nearly equatorial view, see Fig~\ref{fig:aspect_data} (right panel). 
The Asteroid Lightcurve Database (\href{https://www.minorplanet.info/PHP/lcdb.php}{LCDB}) provides the values Amin = 0.06 mag and Amax = 0.13 mag.
}
\label{fig:all_opt_lc}
\end{figure*}

\begin{table*}
\caption{Archived survey images on which the asteroid (78824)~2003~QS13 potentially would have been observed. The limiting magnitude on these images is typically around 18\,mag for 20 sec and 19\,mag for 30 sec exposure times (Exp.) respectively, depending on sky conditions etc. On none of these images the asteroid could be detected as the object was usually too faint (between 20~V-mag and 22~V-mag), with the exception of the date 96-12-17, where the asteroid was at predicted 19.3~V-mag. However, also on those images the asteroid was not found. MJD = JD - 2400000.5 is the Modified Julia Date of the start of the exposure. The <url-prefix> is https://sbnarchive.psi.edu/pds3/neat/geodss} 
\label{tab:archive_images}
\centering
\begin{tabular}{ccccc}
\hline\hline\\[-0.9em]
Image  &  MJD	&  Exp.(s) &  Telescope / Instrument	  &  Datalink\\ 
\hline
961217104449a	&	50434.4477893518  	&	40	&	NEAT-GEODSS-Maui	&	<url-prefix>/g19961217/obsdata/961217104449a.fit.fz \\
961217105739a	&	50434.4567013889	&	40	&	NEAT-GEODSS-Maui	&	<url-prefix>/g19961217/obsdata/961217105739a.fit.fz	\\
961217110944a	&	50434.4650925926	&	40	&	NEAT-GEODSS-Maui	&	<url-prefix>/g19961217/obsdata/961217110944a.fit.fz	\\
970111053115a	&	50459.2300347222	&	40	&	NEAT-GEODSS-Maui	&	<url-prefix>/g19970111/obsdata/970111053115a.fit.fz	\\
970111054321a	&	50459.2384375000	&	40	&	NEAT-GEODSS-Maui	&	<url-prefix>/g19970111/obsdata/970111054321a.fit.fz	\\
970111055852a	&	50459.2492129630	&	40	&	NEAT-GEODSS-Maui	&	<url-prefix>/g19970111/obsdata/970111055852a.fit.fz	\\
980326100223a	&	50898.4183217593	&	20	&	NEAT-GEODSS-Maui	&	<url-prefix>/g19980326/obsdata/980326100223a.fit.fz	\\
980326101659a	&	50898.4284606481	&	20	&	NEAT-GEODSS-Maui	&	<url-prefix>/g19980326/obsdata/980326101659a.fit.fz	\\
980326101920a	&	50898.4300925926	&	20	&	NEAT-GEODSS-Maui	&	<url-prefix>/g19980326/obsdata/980326101920a.fit.fz	\\
980326103127a	&	50898.4385069444	&	20	&	NEAT-GEODSS-Maui	&	<url-prefix>/g19980326/obsdata/980326103127a.fit.fz	\\
980326103353a	&	50898.4401967593	&	20	&	NEAT-GEODSS-Maui	&	<url-prefix>/g19980326/obsdata/980326103353a.fit.fz	\\
980326104827a	&	50898.4503125000	&	20	&	NEAT-GEODSS-Maui	&	<url-prefix>/g19980326/obsdata/980326104827a.fit.fz	\\
980327100349a	&	50899.4193171296	&	20	&	NEAT-GEODSS-Maui	&	<url-prefix>/g19980327/obsdata/980327100349a.fit.fz	\\
980428081456a	&	50931.3437037037	&	20	&	NEAT-GEODSS-Maui	&	<url-prefix>/g19980428/obsdata/980428081456a.fit.fz	\\
980428084448a	&	50931.3644444444	&	20	&	NEAT-GEODSS-Maui	&	<url-prefix>/g19980428/obsdata/980428084448a.fit.fz	\\
980428091458a	&	50931.3853935185	&	20	&	NEAT-GEODSS-Maui	&	<url-prefix>/g19980428/obsdata/980428091458a.fit.fz	\\
980525060228a	&	50958.2517129630	&	20	&	NEAT-GEODSS-Maui	&	<url-prefix>/g19980525/obsdata/980525060228a.fit.fz	\\
980525063130a	&	50958.2718750000	&	20	&	NEAT-GEODSS-Maui	&	<url-prefix>/g19980525/obsdata/980525063130a.fit.fz	\\
980525070207a	&	50958.2931365741	&	20	&	NEAT-GEODSS-Maui	&	<url-prefix>/g19980525/obsdata/980525070207a.fit.fz	\\
\hline
\end{tabular}
\end{table*}

\begin{acknowledgements}
The occultation observations made by Roger Venable (USA), provided by the International Occultation Timing Association (IOTA) and by \citet{herald_asteroid_2019} are acknowledged. Also appreciated are the services and data provided by the Asteroids Dynamic Site (AstDys), JPL Horizon and the Minor Planet Center (MPC). 

This work has made use of data from the European Space Agency (ESA) mission Gaia (https://www.cosmos.esa.int/gaia), processed by the Gaia Data Processing and Analysis Consortium (DPAC, https://www.cosmos.esa.int/web/gaia/dpac/consortium).

The author thanks the referee for the helpful comments and constructive review of the paper. I also like to thank Anna Marciniak (A. Mickiewicz University, Poznan, Poland) for her comments on their work and results on (223) Rosa and  José Luis Ortiz (IAA, Granada, Spain) for reading and commenting the original manuscript.

Finally I like to acknowledge open-source software and tools, like Linux, Julia + Packages, Jupyter, Python + Packages, Gnuplot, GNU Fortran, SQLite, GSHHG (shorelines DB), etc.
\end{acknowledgements}

\bibliographystyle{aa}
\bibliography{mybib}

\end{document}